\documentclass{emulateapj}
\usepackage{graphicx}

\usepackage{amsmath}

\shorttitle{TWO-FLUID MHD SIMULATIONS OF CONVERGING HI FLOWS. II.}
\shortauthors{T. INOUE \& S. INUTSUKA}

\begin{document}

\title{
TWO-FLUID MHD SIMULATIONS OF CONVERGING HI FLOWS IN THE INTERSTELLAR MEDIUM. II: 
ARE MOLECULAR CLOUDS GENERATED DIRECTLY FROM WARM NEUTRAL MEDIUM?
}
\author{Tsuyoshi Inoue\altaffilmark{1}, and Shu-ichiro Inutsuka\altaffilmark{2}}
\altaffiltext{1}{Division of Theoretical Astronomy, National Astronomical Observatory of Japan, Osawa, Mitaka 181-8588 Japan; inouety@th.nao.ac.jp}
\altaffiltext{2}{Department of Physics and Astrophysics, Nagoya University, Chikusa, Nagoya 464-8602, Japan}

\begin{abstract}
Formation of interstellar clouds as a consequence of thermal instability is studied using two-dimensional two-fluid magnetohydrodynamic simulations.
We consider the situation of converging, supersonic flows of warm neutral medium in the interstellar medium that generate a shocked slab of thermally unstable gas in which clouds form.
We found, as speculated in paper I, that in the shocked slab magnetic pressure dominates thermal pressure and the thermal instability grows in the isochorically cooling, thermally unstable slab that leads formation of HI clouds whose number density is typically $n\lesssim 100$ cm$^{-3}$, even if the angle between magnetic field and converging flows is small.
We also found that even if there is a large dispersion of magnetic field, evolution of the shocked slab is essentially determined by the angle between the mean magnetic field and converging flows.
Thus, the direct formation of molecular clouds by piling up warm neutral medium does not seem a typical molecular cloud formation process, unless the direction of supersonic converging flows is biased to the orientation of mean magnetic field by some mechanism.
However, when the angle is small, the HI shell generated as a result of converging flows is massive and possibly evolves into molecular clouds, provided gas in the massive HI shell is piled up again along the magnetic field line.
We expect that another subsequent shock wave can pile up again the gas of the massive shell and produce a larger cloud.
We thus emphasize the importance of multiple episodes of converging flows, as a typical formation process of molecular clouds.
\end{abstract}

\keywords{magnetohydrodynamics --- ISM: general --- method: numerical --- instabilities}

\section{Introduction}
Molecular clouds are the sites of all the present-day star formation.
However, our understanding of the physical conditions of molecular clouds is very limited.
One promising method to reveal the physical conditions of molecular clouds is studying formation of molecular clouds.
As a formation mechanism of molecular clouds, converging, supersonic HI flows that trigger thermal instability to form cold gas clouds have been studied (Ballesteros-Paredes et al. 1999; Hartmann et al. 2001; Inoue \& Inutsuka 2008; Hennebelle et al. 2008; Banerjee et al. 2008; Heitsch et al. 2005, 2006, 2008, 2009; V\'azquez-Semadeni et al. 2006, 2007, see also Koyama \& Inutsuka 2000, 2002).

Recently, using magnetohydrodynamic (MHD) simulations, Hennebelle et al. (2008), Banerjee et al. (2008) and Heitsch et al. (2009) performed the formation of molecular clouds essentially considering a situation where the orientation of the converging flows and mean magnetic field are parallel.
However, in Inoue \& Inutsuka (2008; henceforth paper I), we showed in a two-dimensional two-fluid MHD simulation that, if the converging flows and magnetic field are orthogonal, only the formation of HI clouds ($n\lesssim 100$ cm$^{-3}$) is possible instead of the formation of dense clouds (precursor of molecular clouds; $n\gtrsim 100$ cm$^{-3}$) due to the effect of magnetic pressure.
Using three-dimensional MHD simulation, Heitsch et al. (2009) also obtained results similar to ours even in the case with weak magnetic field ($B=0.5$ $\mu$G).

In the converging flow scenario, to trigger cooling collapse or thermal instability, supersonic flows are necessary (Hennebelle \& P\'erault 1999).
Since magnetic pressure in the diffuse ISM is comparable to thermal pressure, it is natural to assume that the direction of supersonic flows and mean magnetic field has no strong correlation.
Therefore it is important to investigate whether the formation of molecular clouds is possible when the converging flows and mean magnetic field have a small angle.
Using one-dimensional MHD simulation, Hennebelle \& P\'erault (2000) showed that if the initial angle between the magnetic field and the fluid velocity is larger than 20-30 degrees with a few microGauss magnetic field strength, magnetic pressure 
prevents growth of thermal instability.
In multi-dimension, thermal instability can always grow along the magnetic field line (paper I).
But the effect of magnetic pressure drastically change the physical condition of clouds formed by thermal instability, such as density and temperature (paper I).
In this paper, using two-dimensional MHD simulations, we extend paper I and examine the cases where the converging flows and mean magnetic field have angles as Hennebelle \& P\'erault (2000).
This paper is organized as follows.
In \S 2, we provide numerical setup of simulations.
The results of simulations are shown in \S 3.
Finally, in \S 4, we summarize our results and discuss their implications.

\section{Numerical Setup}
\begin{figure}[t]
\epsscale{1.2}
\plotone{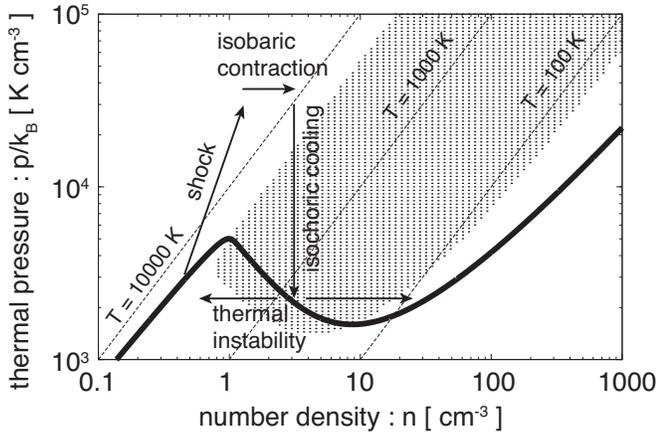}
\caption{
Thermal equilibrium state of our cooling function ({\it thick line}).
The dashed lines represent isothermal lines of $T=10^4,\,10^3$ and $10^2$ K.
In the dotted region, gas is thermally unstable.
The typical gas evolution of Model 1a and 1b is drawn schematically as arrows.
}
\label{f1}
\end{figure}

As in paper I, we consider a weakly ionized interstellar medium using two-fluid approximation incorporating with the effects of coupling between neutral and ionized gases ( friction force/heating, ionization, and recombination), radiative cooling/heating, and thermal conduction.
Details of basic equations (ideal MHD and HD equations with the source terms of two-fluid coupling, heating, cooling, and thermal conduction) and numerical schemes to solve them were developed in paper I.
In Fig. \ref{f1}, the thermal equilibrium state of the cooling function we used is shown as a thick solid line.
We also overlay the dotted regions in which a gas is thermally unstable (see, Balbus 1995; Field 1965; Schwarz et al. 1972; Koyama \& Inutsuka 2002 for linear stability analyses).

We use a two-dimensional, rectangular, numerical domain whose side lengths are $L_{x}=30$ pc and $L_{y}=10$ pc that is divided by $N_{x}\times N_{y}$ uniform square cells ($N_{x}=3\,N_{y}$).
As an initial condition, we set converging flows of warm neutral medium (WNM) that collide head-on at the center of $x$-axis.
Density and thermal pressure of the WNM are $n=0.67$ cm$^{-3}$ and $p_{\rm th}/k_{\rm B}=4,000$ K cm$^{-3}$, which are in thermal equilibrium and typical values of WNM in the ISM.
Our initial condition for the ionized gas is ionization equilibrium to the neutral gas.
Interested readers should refer to Paper I for details.
The velocity field of the converging flows is given such that
\begin{eqnarray}
 v_{x}(x,y)&=&v_{0}(x)+\delta v_{x}(x,y), \\
 v_{y}(x,y)&=&\delta v_{y}(x,y),
\end{eqnarray}
where $v_{0}(x)=v_{\rm cnv}>0$ ($x<L_{x}/2$) and $=-v_{\rm cnv}<0$ ($x>L_{x}/2$)  is the converging component of the velocity, and $\delta v_{x,y}$ is the fluctuation component.
The fluctuation component is given by
\begin{eqnarray}
\delta v_{x} &=& \sum^{10}_{k_{x},k_{y}=0}\,\mathcal{A}(k)\, \cos\left[\frac{2\pi}{L_{y}}\,\left(k_{x}\,x+k_{y}\,y\right)+\phi_{1}(k)\right], \label{vf1}\\
\delta v_{y} &=& \sum^{10}_{k_{x},k_{y}=0}\,\mathcal{A}(k)\, \cos\left[\frac{2\pi}{L_{y}}\,\left(k_{x}\,x+k_{y}\,y\right)+\phi_{2}(k)\right], \label{vf2}
\end{eqnarray}
where $\phi_{1,2}$ are random phases and function $\mathcal{A}$ is chosen so that fluctuations show the Kolmogorov energy spectrum $E(k)\propto k^{-5/3}$ and velocity dispersion of the fluctuations is set at 10\% of the sound speed of the WNM.
Such a spectrum and velocity dispersion are naturally expected in the ISM, since WNM is turbulent with velocity dispersion being transonic at a scale of a few hundreds pc (Dib et al. 2006), which leads velocity dispersion roughly 10 \% of the sound speed at scale of a ten pc.
A uniform magnetic field with an angle of $\theta$ to the $x$-axis is initially imposed.
According to the observations using rotaion/dispersin measures (Rand \& Kulkarni 1989; Beck 2000), the magnetic field strength in the Galactic plane is about 1-5 $\mu$G (approximately equipartition to the thermal energy).
Thus, we use the fixed strength of mean magnetic field $B_{0}= 3.0$ $\mu$G throughout of this paper that gives plasma $\beta$ = 1.5 and meets the above observations.

We also perform simulations with an initially tangled magnetic field, instead of the velocity fluctuations.
In this model, the magnetic field is given by
\begin{eqnarray}
B_{x}(x,y)&=&B_{\rm mean}\,\cos(\theta) + B_{{\rm tangle},x}(x,y),\\
B_{y}(x,y)&=&B_{\rm mean}\,\sin(\theta) + B_{{\rm tangle},y}(x,y),
\end{eqnarray}
where the tangled components of the magnetic field are generated by
\begin{eqnarray}
&&B_{{\rm tangle},x}(x,y)= \partial_{y}\,A_{z}(x,y),\\
&&B_{{\rm tangle},x}(x,y)= - \partial_{x}\,A_{z}(x,y),\\
&&A_{z}(x,y)= \sum^{10}_{k_{x},k_{y}=0}\mathcal{C}(k) \cos\left[\frac{2\pi}{L_{y}}\left(k_{x}x+k_{y}y\right)+\phi(k)\right]. \label{mf1}
\end{eqnarray}
We use random phase as $\phi(k)$.
The spectrum of the magnetic field in the ISM is highly unknown.
Since the spectrum of the magnetic field fluctuation is so unknown, we use $\mathcal{C}(k)\propto k^{-4}$ following Heitsch et al. (2009).
Thus, this model of magnetic field is characterized by the mean magnetic field strength $B_{\rm mean}$, the angle $\theta$, and a root-mean-square of the fluctuating field $\langle B_{\rm tangle} \rangle_{\rm rms}$.

We impose periodic boundary conditions at $y=0$ and $L_{y}$ surfaces.
The fluctuating converging WNM flows are set at $x=0$ and $L_{x}$ surfaces.
The boundary condition of the fluctuation component is given by substituting $x$ for $-v_{0}\,t$ (at $x=0$) and for $L_{x}-v_{0}\,t$ (at $x=L_{x}$) in eqs. (\ref{vf1}), (\ref{vf2}), and (\ref{mf1}).

\begin{deluxetable}{ccccc}
\tablewidth{0pt}
\tablecaption{Model Parameters}
\tablehead{Name &  $v_{\rm cnv}$ [km s$^{-1}$]  & $\theta$ [deg.] & $B_{\rm mean}$ & $\langle B_{\rm tangle} \rangle_{\rm rms}$ }
\startdata
1a & 20.0 & 15.0 & $B_{0}=3.0$ $\mu$G & N/A \\
1b & 20.0 & 40.0 & $B_{0}$ & N/A \\
2a & 10.0 & 15.0 & $B_{0}$ & N/A \\
2b & 10.0 & 40.0 & $B_{0}$ & N/A \\
3a & 20.0 & 15.0 & $B_{0}$ & $B_{0}$ \\
3b & 20.0 & 15.0 & $B_{0}$ & $2.0\,B_{0}$ \\
\enddata
\end{deluxetable}

We examine various situations by changing the free parameters of the models, which are summarized in Table 1.
In this paper, we concentrate on the cases where $\theta$ not equal to 0 and 90 deg., more specifically $\theta$ = 15 and 40 deg., since the case of $\theta$ = 0 is examined in Hennebelle et al. (2008), Banerjee et al. (2008), and Heitsch et al. (2009) and the case of $\theta$ = 90 is examined in paper I and Heitsch et al. (2009).
According to the one-dimensional study by Hennebelle \& P\'erault (2000), the evolution of the shocked medium is affected greatly by the magnetic pressure, when the angle $\theta$ is larger than 20-30 deg. with a few microGauss initial field strength.
Thus, we choose the cases $\theta$ = 15 and 40 deg. that are respectively smaller than and larger than the angle 20-30 deg.

Our initial conditions may correspond to the propagation of shock waves in the ISM, e.g., super-shell due to multiple supernovae and galactic spiral shock.
In general, spiral shocks are highly oblique. However the oblique shock can be transformed to a simple one-dimensional shock by the appropriate Galilean transformation.
Thus, our analyses with various angles of magnetic field lines with respect to the shock normal can also be applied to spiral shocks.
In the case of the super shell, one of the ram pressures of converging flows would be substituted by the thermal pressure of the coronal gas, and in the case of the spiral shock that would be substituted by enhanced thermal pressure due to stellar potential.
In super-shell, after several million years from expansion at which thermal instability is effective in the shell, propagation speed of shock wave is a few times ten km/s depending on the number of SNe (see, e.g., Tomisaka 1998).
Thus, we investigate the cases $v_{\rm cnv} =$ 10 and 20 km/s.

As shown in the following section, the results of most models show formation of clouds whose density is $n\lesssim 100$ cm$^{-3}$.
In such cases the Field length that characterizes the smallest scale of the system is approximately $\gtrsim 0.01$ pc (Inoue et al. 2006).
Thus, in order to resolve the Field length, we use the resolution $N_{x}\times N_{y}=3072\times1024$.
The data of the computations are saved each $0.2$ Myr of the simulations that are used in the analyses given below.

\section{Results}

\subsection{Results of Model 1a and 1b}
\begin{figure}[t]
\epsscale{1.3}
\plotone{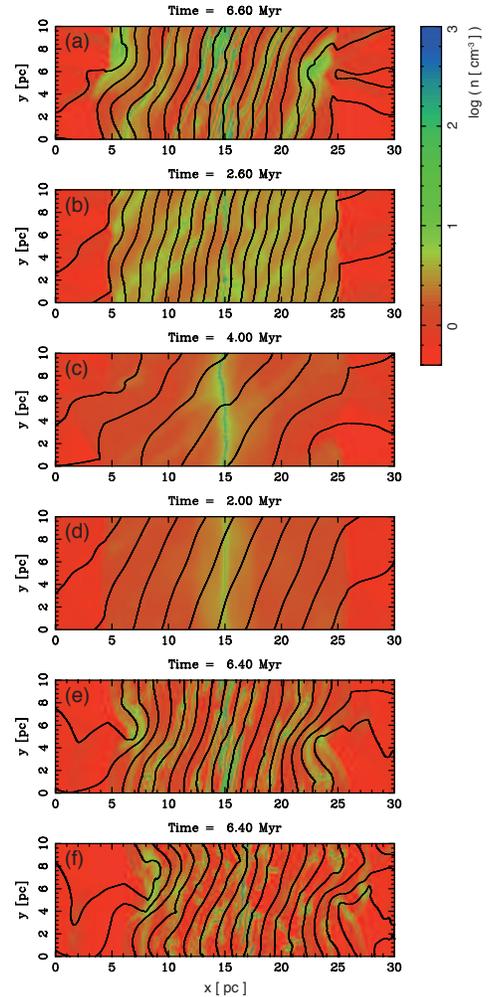}
\caption{
Structures of number densities and magnetic field lines.
Field lines are depicted so that distance from the neighbor field line $l_{m}$ is inversely proportional to the field strength ($B=10\,\mu$G $[l_{m}/1.5\,\mbox{pc}]^{-1}$).
Panels (a)--(f) represent the results of Model 1a, 1b, 2a, 2b, 3a and 3c, respectively.
}
\label{f2}
\end{figure}

We show density structures and magnetic field lines of Model 1a and 1b in Fig. \ref{f2} (a) and (b), respectively.
The depicted times are chosen when the thickness of shocked slab becomes larger than 20 pc.
Here, we define the thickness of the slab such that $l_{\rm slab}=\langle x_{\rm s, right}(y)\rangle_{y}-\langle x_{\rm s, left}(y)\rangle_{y}$, where $\langle \cdots \rangle_{y}$ means average over $y$-direction, $x_{\rm s, left}(y)\equiv {\rm min}[\,x\,|\,B_{\rm y}(x,y)^2/8\pi>3\times10^{-13}\,\mbox{g cm}^{2}\mbox{ s}^{-2}\,]$, and $x_{\rm s, right}(y)\equiv {\rm max}[\,x\,|\,B_{\rm y}(x,y)^2/8\pi>3\times10^{13}\,\mbox{g cm}^{2}\mbox{ s}^{-2}\,]$.
In the two models, structures of shocked slabs are similar.
The evolution of the shocked WNM can be understood as follows (see also paper I):
At first, the WNM  is compressed by the shock by which the WNM kicked out from the thermal equilibrium states and into states dominated by cooling.
Then the slab of shocked cooling gas contracts nearly isobarically in order to oppose the ram pressure of the converging flows.
This contraction leads amplification of magnetic field by compressing transverse components of the magnetic field.
Once the magnetic pressure of the amplified magnetic field becomes comparable to the ram pressure, the isobaric contraction stops and the gas begins to cool nearly isochorically.
The isobaric phase of the evolution does not last long, because the transverse component of the magnetic field is already amplified by the shock compression and the contraction by only a factor of a few can strengthen the magnetic field enough to oppose the ram pressure.
During the isochoric evolution phase, the gas becomes thermally unstable and some portion of the gas, whose density is enhanced by the compressions due to the turbulence given initially, grows toward a stable dense equilibrium, i.e., cold neutral medium (CNM).
Since thermal instability evolves essentially along the magnetic field, filaments of CNM are formed.
On the other hand, the remaining gas continues to be cooled isochorically, until it reaches unstable equilibrium and then grows toward a stable warm phase (WNM).
The sequence of the typical evolutions is drawn schematically in Fig. \ref{f1}.

\begin{figure}[t]
\epsscale{1.2}
\plotone{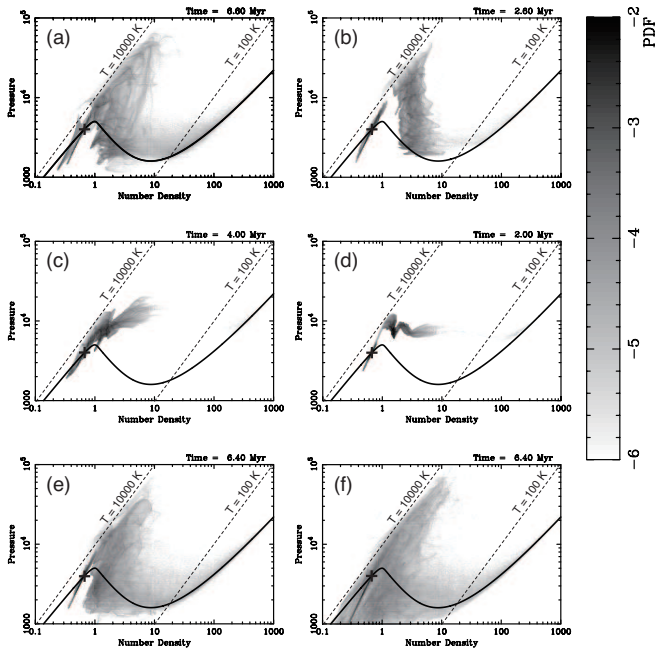}
\caption{
PDFs in $n$--$P$ plane.
Panels (a)--(f) represent the results of Model 1a, 1b, 2a, 2b, 3a and 3c, respectively.
Thermal equilibrium state ({\it solid line}), initial point of simulation ({\it cross}), and isotherms ({\it dashed lines}) are plotted.
}
\label{f3}
\end{figure}

\begin{figure}[t]
\epsscale{1.0}
\plotone{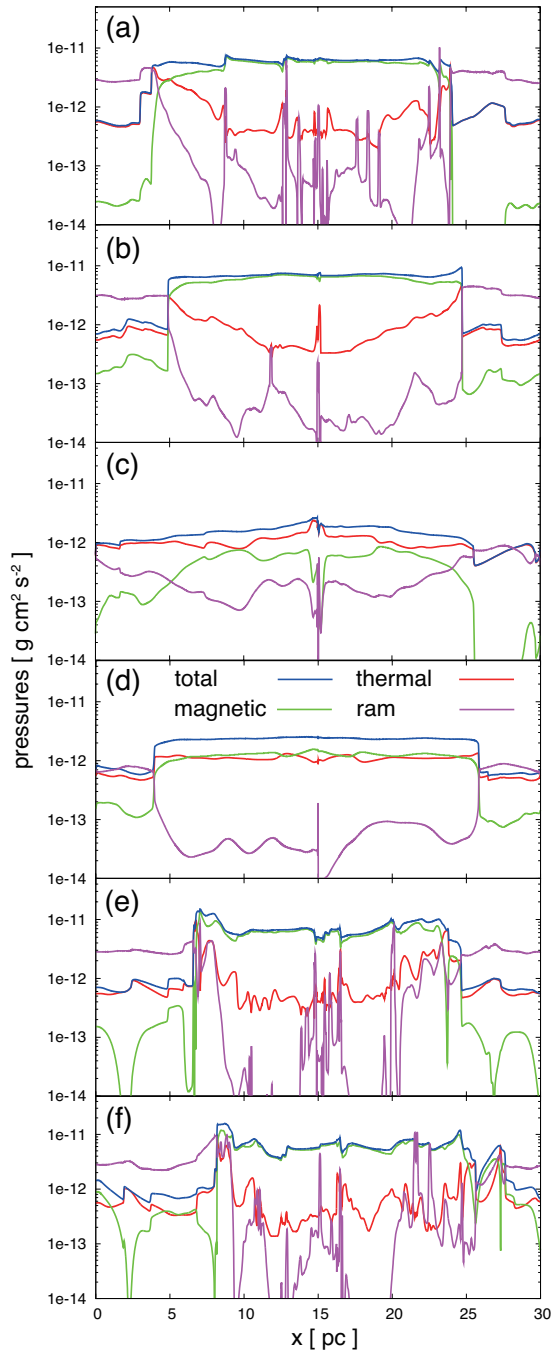}
\caption{
Cross section profiles of thermal ({\it red}), magnetic ({\it green}), total ({\it blue}) and ram ({\it magenta}) pressures at $y=5.0$ pc.
Panels (a)--(f) represent the results of Model 1a, 1b, 2a, 2b, 3a and 3c, respectively.
}
\label{f4}
\end{figure}

In Fig. \ref{f3} (a) and (b), we plot the probability distribution function (PDF) of the gas that confirms the evolution discussed above.
Note that since the timescale of thermal instability toward the WNM phase is approximately 5 Myr (Inoue et al. 2006; 2007), which is longer than the timescale of the instability toward the CNM ($\lesssim 1$ Myr), the diffuse gas component in the shocked slab is still growing toward the WNM phase.
In Fig. \ref{f4} (a) and (b), we plot cross section profiles of the thermal ($p_{\rm th}$), magnetic ($B_{y}^2/8\pi$), total ($p_{\rm th}+B_{y}^2/8\pi$), and ram pressure ($\rho v_{x}^2$) at $y=5.0$ pc.
We can confirm that the shocked slab is supported by the magnetic pressure.
Note that From the instant PDF in Fig. \ref{f3} and cross section in Fig. \ref{f4}, we can also know the information about time evolution of shocked gas.
The gas in the downstream at the distance $r$ from the shock front roughly shows the state of the gas at the time $r/v_d$ since 
shocked, where $v_d$ is the downstream velocity at the shock rest frame.
From this, we can also see that the magnetic pressure promptly dominates the thermal pressure inside the slab that corresponds to the briefness of the isochoric contraction phase.

Densities of the CNMs are typically $\lesssim 100$ cm$^{-3}$, and temperatures are $\sim 50$ K corresponding to HI clouds.
High density clumps of $n>100$ cm$^{-3}$ are formed only at an early stage around the center of the numerical domain ($x\simeq 15$ pc), since high density CNMs can be generated until the thermal pressure of the slab begins to decrease, i.e., until the magnetic pressure dominates the thermal pressure.
To see this, we plot the average column density evolutions of the high density ($n>200$ cm$^{-3}$) and low density ($200$  cm$^{-3}$$>n>20$  cm$^{-3}$) CNMs along the $x$-axis in Fig. \ref{f5} (a).

Note that only the magnetic field amplification a few times the initial strength makes the opposition of the magnetic pressure and the ram pressure possible.
Thus, the magnetic field strength in the slab is not so high in the simulations ($B\sim 10$ $\mu$G) that will decrease to average field strength in the ISM once the converging flows stop.

\subsection{Results of Model 2a and 2b}
We show density structures and magnetic field lines of Model 2a and 2b in Fig. \ref{f2} (c) and (d), respectively.
As in Model 1a and 1b, the depicted times are chosen when the thickness of the shocked slab becomes larger than 20 pc.
Since the velocity of the converging flows is smaller, the shock waves propagate faster than in Model 1a and 1b.

The two models are apparently similar.
However, details of structures and their evolutions are different.
In the case of Model 2a four shock waves are induced, i.e., fast and slow shocks propagate to $\pm$ x directions, while slow shocks are not clearly seen in Model 2b like Model 1a and 1b.
This difference is seen more clearly in earlier stages.
In Fig. \ref{f6}, we show density structures and magnetic field lines of Model 2a and 2b when the thickness of the shocked slabs is $\sim 10$ pc.
In general, fast shocks amplify the magnetic field by compressing field lines, while slow shocks lower field strength by stretching the field lines, i.e., in the case of fast shock, the angle between magnetic field and shock normal becomes larger in downstream compared to the upstream, and in the case of slow shock, the angle becomes smaller.
By using these basic characteristics of MHD shocks, we can identify in the top panel of Fig. 6 (Model 2a) that there are fast shocks at $x \simeq$ 10 and 20 pc and there are slow shocks at $x \simeq$ 12.5 and 17.5 pc.
On the other hand, in the bottom panel of Fig. 6 (Model 2b), there are only two fast shocks.

\begin{figure}[t]
\epsscale{1.}
\plotone{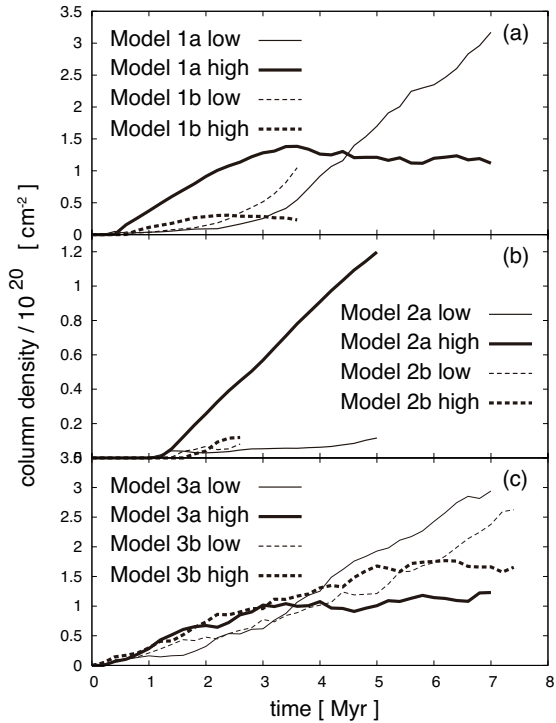}
\caption{
Average column density evolutions of the high density ($n> 200$ cm$^{-3}$) and low density ($200$ cm$^{-3}$ $> n > 20$  cm$^{-3}$) CNMs along the $x$-axis.
Panels (a)--(c) represent the results of Model 1a, 1b, 2a, 2b, 3a, and 3b, respectively.
Thick (thin) lines show the column density evolutions of high (low) density CNMs.
}
\label{f5}
\end{figure}

The structures of the pressures in the two models also differ.
Fig. \ref{f4} (c) and (d) shows the cross section profiles of the thermal, magnetic, total (thermal plus magnetic), and ram pressures at $y=5.0$ pc.
In Model 2a, owing to the slow shocks magnetic pressure does not dominate thermal pressure in the shocked slab.
In this case, the condensation driven by cooling can proceed without decreasing thermal pressure greatly (nearly isobaric contraction) that leads formation of dense CNM whose number density is larger than 200 cm$^{-3}$.
This evolution is similar to the evolution of shocked gas in unmagnetized media (see, paper I).
In the case of Model 2b, on the other hand, the magnetic and thermal pressures are comparable (see Fig. \ref{f4} [d]).

The magnetic pressures, in particular those in Model 2b, are still increasing to dominate the thermal pressure.
Since the shock propagation speed is fast, the shock waves reach computational boundaries within a few Myr that makes the timescale short within which we can examine the evolution of the shocked slab.
Thus, we have not confirmed the clear domination of the magnetic pressure in the slab.
However, in Model 2b, we expect from the absence of the slow shocks that the shocked slab would evolve into a two-phase medium composed of low-density CNM ($n<200$ cm$^{-3}$) and WNM as in Model 1a and 1b.
In order to confirm the domination of magnetic pressure, simulation with a much larger numerical domain is necessary.

We can also see the difference between the two models in Fig. \ref{f5} (b).
In Model 2a, the growth of column density of high density CNM does not show saturation unlike Model 1a and 1b, while Model 2b evolves like Model 1a and 1b, although the timescale we have examined is short.
Note that the PDFs in $n$-$P$ plane of Model 2a and 2b plotted respectively in Fig. \ref{f3} (c) and (d) does not clearly differ between the two models.
This is because the slab of Model 2b is still in the isobaric contraction phase.
The difference would be clear once the magnetic pressure dominates the thermal pressure in Model 2b that leads the evolution of the isochoric cooling in n-P plane.

\begin{figure}[t]
\epsscale{1.0}
\plotone{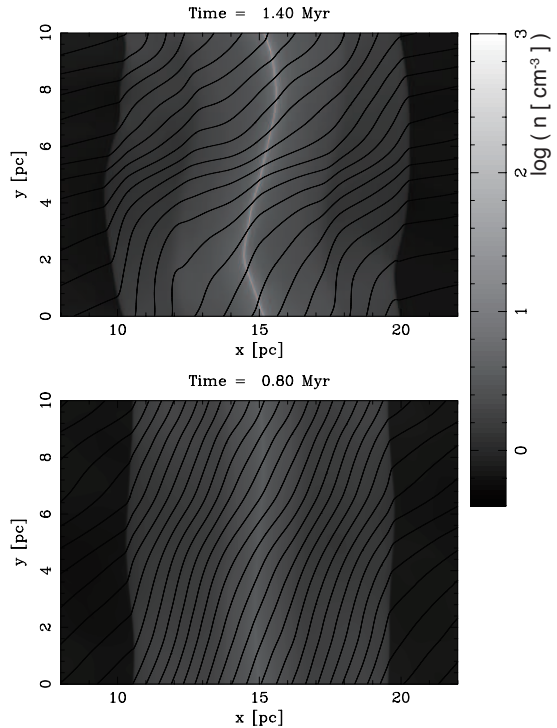}
\caption{
Structures of number densities and magnetic field lines when the thickness of the shocked slabs is $\sim 10$ pc.
Field lines are depicted six times thicker than Fig. \ref{f2}.
Top and bottom panels show the results of Model 2a and 2b, respectively.
}
\label{f6}
\end{figure}

What controls bifurcation of the number of induced shocks?
The condition of the bifurcation can be derived from the MHD Rankine-Hugoniot relations:
\begin{eqnarray}
s\,(\rho_{0}-\rho_{1})&=&\rho_{0}\,v_{x,0}-\rho_{1}\,v_{x,1}, \label{RK1}\\
s\,(\rho_{0}\,v_{y,0}-\rho_{1}\,v_{y,1})&=&\left( \rho_{0}\,v_{x,0}\,v_{y,0}-\frac{B_{x}\,B_{y,0}}{4\,\pi}\right)\nonumber\\
&&-\left( \rho_{1}\,v_{x,1}\,v_{y,1}-\frac{B_{x}\,B_{y,1}}{4\,\pi}\right),\label{RK2}\\
s\,(B_{y,0}-B_{y,1})&=&(v_{x,0}\,B_{y,0}-v_{y,0}\,B_{x})\nonumber\\
&&-(v_{x,1}\,B_{y,1}-v_{y,1}\,B_{x}), \label{RK3}
\end{eqnarray}
where the subscripts 0 and 1 respectively denote the upstream and downstream value, $s$ is the shock propagation speed, and we have used the fact that $B_{x}=\,$constant across the shock wave owing to $\vec{\nabla}\cdot\vec{B}=0$.
The equations (1)--(3) represent the conservation laws of mass, tangential component of momentum, and tangential component of electric field, respectively.

\begin{figure}[t]
\epsscale{1.2}
\plotone{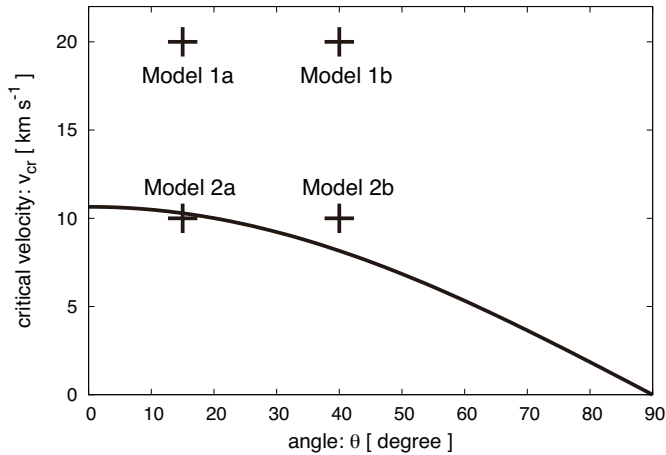}
\caption{
Critical velocity as a function of $\theta$ with the typical parameters $B=3.0\,\mu$G, $n_{0}=0.67$ cm$^{-3}$, and $r=4$.
Conditions of Model 1a, 1b, 2a, and 2b are plotted as crosses.
}
\label{f7}
\end{figure}

Immediately after the collision of the converging flows, $v_{x,1}$ and $v_{y,0}$ should be zero.
By using these conditions and eqs. (\ref{RK1})--(\ref{RK3}), eliminating $s$ and $v_{y,1}$, we obtain
\begin{equation}
\frac{(\rho_{1}-\rho_{0})\,B_{x}^{2}}{\rho_{0}\,\rho_{1}\,v_{x,0}}-\frac{\rho_{0}\,v_{x,0}}{\rho_{1}-\rho_{0}}=\frac{v_{x,0}\,B_{y,0}}{(B_{y,0}-B_{y,1})}.
\end{equation}
In the case of the fast shock, if $B_{y,0}>0$ ($<0$), $B_{y,0}-B_{y,1}$ should be less (greater) than zero.
Therefore, collision of the converging flow induces only the fast shock, when
\begin{equation}
v_{\rm cnv}=v_{x,0}>\frac{r-1}{\sqrt{r}}\,\frac{B\,\cos{\theta}}{\sqrt{4\pi\,\rho_{0}}}\equiv v_{\rm cr},\label{VC}
\end{equation}
where we have introduced the compression ratio of the density $r\equiv \rho_{1}/\rho_{0}$, and $\theta$ is the angle between converging flows and upstream magnetic field.
If the velocity of the converging flows is smaller than the critical velocity $v_{\rm cnv}<v_{\rm cr}$, additional slow shocks are necessary in order to obtain physical down stream state.
In Fig. \ref{f7}, we show the critical velocity as a function of $\theta$ by using the typical parameters $B=3.0\,\mu$G, $n_{0}=0.67$ cm$^{-3}$ and $r=4$.
We also plot the conditions corresponding to Model 1a, 1b, 2a, and 2b.

In order to form dense CNM (precursor of molecular clouds) whose number density is larger than 200 cm$^{-3}$, the shocked gas should condense without decreasing thermal pressure, i.e., the magnetic pressure should not dominate in the slab.
This requires the slow shocks that lower magnetic field strength in the slab.
Thus, if the velocity of converging flows is smaller than the critical velocity a dense CNM sheet is formed as Model 2a.
On the other hand, if the velocity of converging flows is larger than the critical velocity, the magnetic pressure dominate in the slab that leads isochoric cooling and formation of low-density HI clouds ($n<100$ cm$^{-3}$) as Model 1a, 1b, and as expected in Model 2b.

Therefore, only the case where $v_{\rm cnv} < v_{\rm cr}$ (Model 2a) can generate molecular clouds directly from WNM.  However, the column density of dense CNM sheet formed in Model 2a (the precursor of molecular cloud) is at least an order of magnitude smaller than that of molecular clouds ($>10^{21}$ cm$^{-2}$).
Thus, in order to form molecular clouds, the converging flows have to last a long time.
Note that, even in such case, the shocks continue to propagate outwards making direct formation of molecular clouds difficult as follows:
The shocked gas that accretes to the dense CNM sheet is provided with sub-slow speed far from the dense CNM sheet.
In the ISM, such a slow flow is easily affected by, e.g., SNe and stellar winds by which flow direction would change.
As a result, the accretion to the dense CNM is stopped leading to failure of molecular cloud formation.

\subsection{Results of Model 3a and 3b}
In Model 3a and 3b, instead of the velocity fluctuations, we impose the highly tangled magnetic field.
Fig. \ref{f2} (e) and (f) respectively show density structures and magnetic field lines of the results of Model 3a and 3b at which the thickness of the shocked slab becomes larger than 20 pc\footnote{In Model 3b, we define $x_{\rm s, left (right)}(y)\equiv {\rm min (max)}[\,x\,|\,B_{\rm y}(x,y)^2/8\pi>3\times10^{-12}\,\mbox{g cm}^{2}\mbox{ s}^{-2}\,]$, since the initial transverse component of magnetic field in Model 3b is larger than the other models.}.
We see that the results are close to that of Model 1a.
The PDFs in Fig. \ref{f3} (e) and (f), the cross-section profiles in Fig. \ref{f4} (e) and (f), and the average column density evolutions in Fig. \ref{f5} (c) also show behavior similar to those of Model 1a and 1b.
Since the condition in Model 3b is more turbulent than Model 3a due to the larger tangled component of magnetic field, the distribution of the gas in n-P plane in Model 3b is more dispersed than Model 3a.

From these results, we conclude that evolution of the shocked slab is essentially determined by the angle between the mean magnetic field and converging flows, even if the mean magnetic field is weaker than the tangled component.
The reason why the mean magnetic field determines the evolution of the shocked slab is that the tangled components of the magnetic field behave as Alfv\'en waves which eventually decay due to the cascade of MHD turbulence within Alfv\'en crossing time (Goldreich \& Sridar 1995; Cho \& Lazarian 2005).
Therefore, if $v_{cnv}>v_{cr}$, even when the tangled components dominate the mean field, the fate of the shocked slab is determined by the mean magnetic field.

One may wonder the different conclusion given in Heitsch et al. (2009) in which tangled magnetic field component affect the post shock structure greatly.
In Heitsch et al (2009), they compared the models in which the mean magnetic field is parallel to the flow direction  except their Model Y05.
In these cases, the shocked gas evolves in different ways depending on the local orientation of magnetic field as follows:  If the local magnetic field is parallel to the converging flows, the magnetic field is not enhanced that would leads formation of dense CNM.
On the other hand, if local magnetic field is perpendicular to the flow, the enhanced magnetic pressure would prevents formation of dense CNM that leads void.
Thus, the tangled component of magnetic field can change the morphology of the clouds, if we consider the case of vanishing transverse mean magnetic field.
In our calculation, we consider the models in which the mean magnetic field and the converging flows have finite angles.
In these cases, if $v_{cnv} > v_{cr}$, the magnetic pressure eventually dominate the thermal pressure in the post shock media that prevents direct formation of dense CNM and leaves two-phase media composed of low-density CNM and WNM.

\section{Summary and Discussion}
\subsection{Summary}
In this paper, using two-dimensional two-fluid MHD simulations, we studied the formation of interstellar clouds as a consequence of thermal instability in the shocked slab generated by converging WNM flows.
Our findings are as follows:
\begin{itemize}
\item The evolution of the shocked slab is controlled by the parameters $v_{\rm cnv}$ (the velocity of the converging flows) and $\theta$ (the angle between the converging flows and the mean magnetic field).

\item If the velocity of the converging flows is larger than the critical velocity $v_{\rm cr}$ derived in eq. (\ref{VC}) and $\theta$ is at least larger than 15 deg., HI clouds embedded in WNM are formed in the slab, as we speculated in paper I.
In general, in the case with a moderate magnetic field strength ($B=3$ $\mu$G), the critical velocity $v_{\rm cr}$ is comparable to  the sound speed of the WNM $\sim$ 10 km s$^{-1}$ (see, Fig. \ref{f4}) depending on $\theta$.
It is known that in order to trigger cooling collapse or thermal instability, supersonic flows are necessary (Hennebelle \& P\'erault 1999).
Therefore, this would be the most  probable fate of the converging flows in the ISM.

\item If the velocity of the converging flows is smaller than the critical velocity or the angle $\theta$ is nearly zero, dense clouds that possibly evolve into molecular clouds can be formed in the slab (see, Hennebelle et al. 2008; Banerjee et al. 2008; Heitsch et al. 2009 in the case of $\theta=0$).

\item The entanglement of the magnetic field is not an important control parameter for the evolution of the shocked slab.
The important parameter is the angle between {\it mean magnetic field} and the converging flows.
Even if strength of the mean magnetic field is weaker than the strength of the entangled component (dispersion), the mean magnetic field determines evolution of the shocked slab.
\end{itemize}

In summary, the direct formation of molecular clouds from WNM by converging flows is possible, only if the angle between the mean magnetic field and the converging flows is close to zero or the velocity of converging flows is less than the critical velocity given by eq. \ref{VC}. 

Note that our results are obtained using two-dimensional simulations that may change the configuration of clouds in three dimensions due to the modes of fluid motion that are not appear in planar case, e.g., the interchange mode.
Thus, 3D simulation should be performed in order to confirm our results. 
However, the model Y05 ($\theta=90$ deg. and $B=0.5$ $\mu$G) of Heitsch et al. (2009) shows that, even when the initial magnetic field strength is an order of magnitude weaker than the equipartition, the magnetic pressure dominate thermal pressure in the shocked slab.
This suggests that there is no mechanism that reduces the enhanced magnetic pressure even in 3D.
Thus, we can expect that the enhancement of magnetic pressure makes the direct formation of molecular cloud difficult even in 3D.

In this paper, we also neglect the effect of self-gravity. As discussed in Koyama \& Inutsuka (2002), the scale of the clumps of CNM generated by thermal instability is much smaller than the Jeans length.
Thus, we can neglect the effect of self-gravity on each individual CNM clumps.
Furthermore, the mean densities in the shocked slabs are smaller than 10 cm$^{-3}$ (see Fig. 8 below) indicating the Jeans length of shocked slab larger than 50 pc and the free-fall time larger than 50 Myr, even if we underestimate the mean temperature of the shocked slabs as 100 K.
Thus, we can safely neglect the effect of self-gravity that leads contraction of shocked slab to form molecular cloud.
Note that this is not the case when theta = 0 deg, since the mean density of the slab would be much larger.

\subsection{Discussion}
We have shown that the direct formation of molecular clouds from WNM seems difficult unless some mechanisms creates biased supersonic flows of WNM that is parallel to the mean magnetic field orientation.
Our results would be also applied to the situations of propagation of single shock wave, e.g., super-shell due to multiple supernovae (SNe) and spiral shock wave.
In the cases of the super shell, one of the ram pressures of converging flows that bound the shell would be substituted by the thermal pressure of the hot gas, and in the case of the spiral shock that would be substituted by enhanced thermal pressure that balances the stellar potential.

However, if we consider some mechanisms that further pile up the gas in the two-phase shell generated by the converging flows, the formation of molecular clouds is possible.
For example, in the case of Model 1a, the mean density of the gas in the two-phase shell is $\sim 10$ cm$^{-3}$ that is more than ten times higher than WNM.
Thus, molecular cloud with $A_{V}\sim 1$ mag. would be formed if the two-phase gas in the shell is piled up again only over tens of parsec along the magnetic field line.
The ISM is known to be frequently (once in a million years) swept up by SN shocks, which propagate a few hundred parsecs until decay.
Thus, it can induce formation of molecular clouds by piling up the massive two-phase shell.
As in the case of the converging flows studied in this paper, SN shock can pile up gas only when it propagates along the magnetic field line in the two-phase shell.
Despite this, it is promising because of the high event rate of SN.
We note the difficulty in propagation of a single SN shock wave piling up enough WNM to generate molecular clouds even if it propagates along magnetic field line because of its short lifetime.
The single SN shock can generate molecular clouds only when the density of preshock gas is enhanced by something just as we consider here.

\begin{figure}[t]
\epsscale{1.2}
\plotone{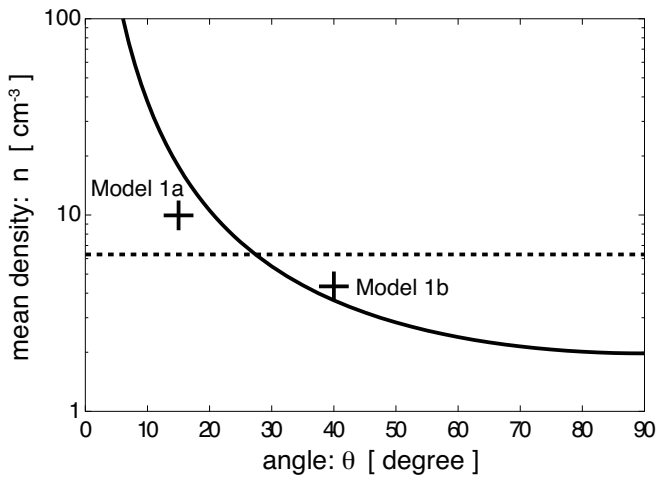}
\caption{
Solid line shows the theoretical post shock mean density derived in equation (16).
The parameters $\rho_{0}$ and $v_{x,0}$ are the same to Model 1a and 1b and we use $B_{y,0}(\theta)=B \cos(\theta)$, where $B=3$ $\mu$G.
The post shock values $n_1$ and $B_{y,1}$ are calculated by solving MHD shock jump conditions using the preshock conditions of Model 1a and 1b.
The mean densities of the results of Model 1a and 1b are also plotted as crosses, which are evaluated at the times depicted in Fig. \ref{f2}.
The dotted line represents the critical mean density [eq. (15)].
If the mean density is larger than the critical value, a single SN shock can generate molecular cloud.
}
\label{f8}
\end{figure}

What is the angle $\theta$ below which the converging flows form the massive shell that can evolve into molecular clouds?
If we assume an SN shock wave that pile up again the two-phase shell, the mean density in the two-phase shell should be larger than
\begin{equation}\label{CD}
\langle n\rangle \gtrsim 6.3 \mbox{ cm}^{-3} \left( \frac{l_{\rm SN}}{10^2\mbox{ pc}} \right)^{-1}\,\left( \frac{A_{V}}{1\mbox{ mag.}} \right),
\end{equation}
where $l_{\rm SN}$ is the propagation length of SN shock until decay and $A_{V}$ is the visual extinction of the formed molecular cloud.
In \S 3.1, we have shown that the mean density enhancement due to the isobaric contraction of the shocked gas is stopped when the ram pressure of the converging flows and the magnetic pressure in the shell are balanced.
This indicates that the isobaric contraction enhances the post shock gas by a factor of $(\rho_{0} v_{x,0}^{2})/(B_{y,1}^{2}/8 \pi)$, where subscript 0 stands for preshock state (WNM) and subscript 1 stands for postshock state (immediately behind 
the shock front).
Note that $B_{y,1}$ depends on $\theta$ (the angle between upstream magnetic field and shock normal).
Thus, the mean density of the shocked slab (two-phase shell) can be estimated by
\begin{equation}\label{MD}
\langle n\rangle \sim n_{1}\,\frac{\rho_{0}\,v_{x,0}^{2}}{B_{y,1}^{2}(\theta)/8\pi},
\end{equation}
where subscript $0$ stands for preshock state (WNM), subscript $1$ stands for postshock state (immediately behind the shock wave), the $x$- and $y$-axes respectively represent directions normal to and tangential to the shock front, and $\theta$ is the angle between the preshock magnetic field and flow.
In Fig. \ref{f8}, we plot the mean density $\langle n \rangle$ (eq. \ref{MD}) against the angle between the converging flows and mean magnetic field $\theta$ by using preshock conditions of Model 1a and 1b.
The postshock condition is calculated by solving the MHD Rankine-Hugoniot relation.
We also plot the mean density of our Model 1a and 1b as crosses, which are evaluated at the times depicted in Fig. \ref{f2}, and plot the estimate of the critical mean density in equation (\ref{CD}) as a dotted line.
If the mean density is larger than the critical density, a SN shock wave can generate molecular cloud by piling up the shell.
The results of simulations agree well with theoretical estimation of equation (\ref{MD}).
From this figure, we expect that the formation of molecular clouds by piling up the two-phase shell is possible roughly when $\theta\lesssim 30$ deg..

We conclude that the formation scenario of molecular clouds from the two-phase medium (HI clouds embedded in WNM) is expected to be a typical scenario.
In this scenario, the fundamental building blocks of molecular clouds are not WNM but H I clouds consistent with the suggestions from observations of molecular clouds in Local Group galaxies (Fukui 2007; Blitz et al. 2007).
Recently, we showed that a very strong shock wave that sweeps up two-phase gas cause turbulence and magnetic field amplification in the post shock medium (Inoue et al. 2009, see also Giacalone \& Jokipii 2007).
A similar phenomenon would also affect the physical conditions of molecular clouds formed from HI clouds  as a result of shock sweeping that will be examined in a future paper.

Note that the above speculation is obtained by assuming that the mean magnetic field strength in WNM is near equipartition to the thermal energy ($B \sim 3$ $\mu$G).
If the mean magnetic field strength is much larger than the equipartition, the direct formation of molecular clouds becomes easier, since the critical velocity derived in eq. (14) depends linearly on $B$.
Also, the weak magnetic field would lead the direct formation of molecular clouds, if magnetic pressure is too weak to stop the isobaric contraction of the shocked slab.

\acknowledgments
Numerical computations were carried out on NEC SX-9 and XT4 at the Center for Computational Astrophysics (CfCA) of National Astronomical Observatory of Japan.
This work is supported by a Grant-in-aid from the Ministry of Education, Culture, Sports, Science, and Technology (MEXT) of Japan, No. 21740146 (T.I.), and No.15740118, No.16077202 and No.18540238 (S.I.).

\end{document}